\documentclass[%
preprint,
 amsmath,amssymb,
aps,
prb,
]{revtex4-1}
\usepackage{graphicx}
\usepackage{dcolumn}
\usepackage{bm}

\newcommand{\phext}{\ensuremath{\varphi_\mathrm{ext}}}
\newcommand{\taua}{\ensuremath{\tau_\mathrm{avg}}}
\newcommand{\hebt}{\ensuremath{H_{\mathrm{EB}}(\varphi_{\mathrm{ext}})}}
\newcommand{\hct}{\ensuremath{H_\mathrm{C}(\varphi_{\mathrm{ext}})}}
\newcommand{\heb}{\ensuremath{H_\mathrm{EB}}}

\newcommand{\hc}{\ensuremath{H_\mathrm{C}}}

\newcommand{\hcmin}{\ensuremath{H_\mathrm{C,min}}}
\newcommand{\kf}{\ensuremath{K_\mathrm{F}}}
\newcommand{\tf}{\ensuremath{t_\mathrm{F}}}
\newcommand{\irmn}{Ir$_{17}$Mn$_{83}$}
\newcommand{\cofe}{Co$_{70}$Fe$_{30}$}
\newcommand{\betf}{\ensuremath{\beta_{\mathrm{F}}}}

\newcommand{\aleb}{\ensuremath{\gamma_{\mathrm{EB}}}}
\newcommand{\alra}{\ensuremath{\gamma_{\mathrm{RMA}}}}
\newcommand{\alf}{\ensuremath{\gamma_{\mathrm{F}}}}
\newcommand{\jadd}{\ensuremath{J_{\mathrm{add}}^{\mathrm{eff}}}}
\newcommand{\msat}{\ensuremath{M_{\mathrm{sat}}}}
\newcommand{\hext}{\ensuremath{H_{\mathrm{ext}}}}
\newcommand{\vhext}{\ensuremath{\vec{H}_{\mathrm{ext}}}}
\newcommand{\gra}{\ensuremath{^{\circ}}}

\newcommand{\jeb}{\ensuremath{J_\mathrm{EB}^\mathrm{eff}}}
\newcommand{\jhc}{\ensuremath{J_\mathrm{C}^\mathrm{eff}}}

\begin{document}


\title{Grain size correlated rotatable magnetic anisotropy in polycrystalline exchange bias systems}

\author{Nicolas David M{\"u}glich}
\email[]{nicolas.mueglich@physik.uni-kassel.de}
\affiliation{Institute of Physics and Center for Interdisciplinary Nanostructure Science and Technology (CINSaT),
University of Kassel, Heinrich-Plett-Strasse 40, 34132 Kassel, Germany}

\author{Gerhard G{\"o}tz}

\affiliation{Center for Spinelectronic Materials and Devices, Physics Department, Bielefeld University, Universit{\"a}tsstra{\ss}e 25, 33615 Bielefeld, Germany}

\author{Alexander Gaul}
\author{Markus Meyl}

\affiliation{Institute of Physics and Center for Interdisciplinary Nanostructure Science and Technology (CINSaT),
University of Kassel, Heinrich-Plett-Strasse 40, 34132 Kassel, Germany}


\author{G{\"u}nter Reiss}
\author{Timo Kuschel}

\affiliation{Center for Spinelectronic Materials and Devices, Physics Department, Bielefeld University, Universit{\"a}tsstra{\ss}e 25, 33615 Bielefeld, Germany}

\author{Arno Ehresmann}
\affiliation{Institute of Physics and Center for Interdisciplinary Nanostructure Science and Technology (CINSaT),
University of Kassel, Heinrich-Plett-Strasse 40, 34132 Kassel, Germany}

\date{\today}

\begin{abstract}
Angular resolved measurements of the exchange bias field and the coercive field are a powerful tool to distinguish between different competing magnetic anisotropies in polycrystalline exchange bias layer systems. No simple analytical model is as yet available, which considers time dependent effects like enhanced coercivity arising from the grain size distribution of the antiferromagnet.\\
In this work we expand an existing model class describing polycrystalline exchange bias systems by a rotatable magnetic anisotropy term to describe grain size correlated effects. Additionally, we performed angular resolved magnetization curve measurements using vectorial magnetooptic Kerr magnetometry. Comparison of the experimental data with the proposed model shows excellent agreement and reveals the ferromagnetic anisotropy and properties connected to the grain size distribution of the antiferromagnet. Therefore, a distinction between the different influences on coercivity and anisotropy becomes available.
\end{abstract}


\maketitle


\section{Introduction}
\begin{figure}[htbp]
\includegraphics{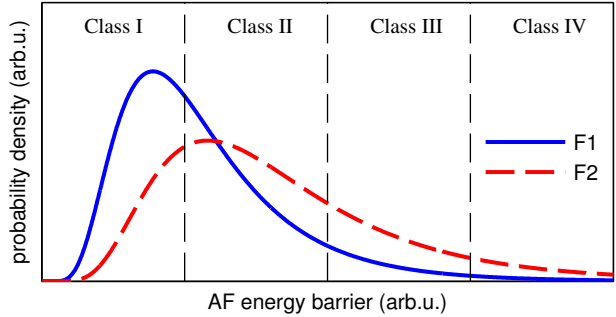}
\caption{\label{fig:grainsizes}Schematic probability distribution of energy barriers in AF grains. F1 and F2 are lognormal functions related to systems with a different average energy barrier. AF grains may be divided into four categories (see text) depending on their energy barriers.}
\end{figure}

\begin{figure}[htbp]
\includegraphics{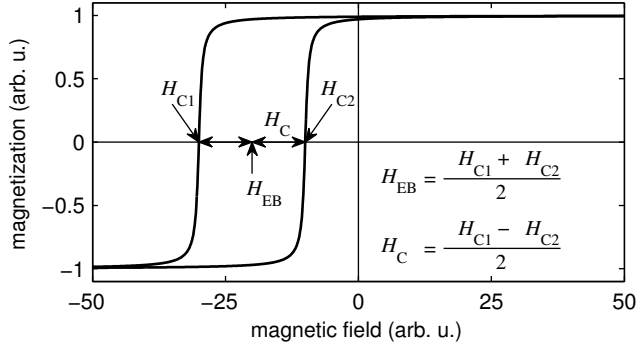}
\caption{\label{fig:hys_schema}Schematic view of a magnetization curve of an EB layer system. The characteristic quantities exchange bias field \heb~(the curve shift) and coercive field \hc~(the broadness of the hysteretic curve) are calculated with the two coercive fields $H_\text{C1}$ and $H_\text{C2}$, at which the magnetization along the measurement direction equals zero.}
\end{figure}

Exchange bias (EB), firstly discovered by Meiklejohn and Bean in 1956\cite{Mei56_mod,mei57_new}, is a magnetic interface effect between a ferromagnetic (F) and an antiferromagnetic (AF) layer\cite{Mei56_mod}. The exchange interaction between individual magnetic moments of the F and the AF across the interface leads to unidirectional\cite{Mei56_mod} and rotatable magnetic anisotropy\cite{Ges02_rot,Kim03_rotaniso,Rad08_rev} in the F. EB is nowadays widely used in magnetic sensor heads to pin the magnetization direction of the magnetic reference electrode\cite{Nog99_rev}. Recently, new interest in an improved description of the EB effect arose\cite{Ehr15_sensor} as ion bombardment induced magnetic patterning of EB layer systems allows tailoring artificial magnetic stray field landscapes\cite{Ehr06_pattern} which are very likely to be a central part of magnetic particle transport in lab-on-chip applications for, e.g., biosensing\cite{ehresmann2011asymmetric,holzinger2015directed,Ehr15_sensor}.\\
Although the EB effect was investigated for almost 60 years, a complete theoretical description is still not available. Recently, a promising model\cite{OGr10_york} for polycrystalline magnetic thin films was developed, which detailed similar earlier ideas\cite{soeya1996nio,Ehr05_mod}. This model takes into account the granular structure of a polycrystalline AF\cite{Ful72_mod}, dividing grains into categories\cite{soeya1996nio} based on their individual relaxation times
\begin{equation}
\tau_\mathrm{i}=\frac{1}{f_0} \exp \left[-\frac{\Delta E}{k_\mathrm{B}T}\right],\Delta E=K_{\mathrm{AF,i}}V_{\mathrm{AF,i}}.
\label{relaxtime}
\end{equation}
Here, $f_0$ is the characteristic frequency for spin reversal\cite{Val01_attempt}, $T$ is the temperature and $k_\mathrm{B}$ is Boltzmann's constant. $\Delta E$ is the energy barrier between a local and a global energy minimum in the potential energy landscape as a function of angle between F and pinned AF moment\cite{Ehr05_mod}. In first order, it can be written as the product of magnetic anisotropy $K_{\mathrm{AF,i}}$ and volume $V_{\mathrm{AF,i}}$ of the respective grains. Polycrystalline layer systems typically show a distribution of grain sizes\cite{Val08_grains}, resulting in a distribution of relaxation times over several orders of magnitude\cite{OGr10_york}.
\\Thermal stability of grains may be defined by comparing the relaxation time to the measurement conditions\cite{OGr10_york}, leading to a classification of grains into categories\cite{soeya1996nio,Ehr11_drift} as shown in figure \ref{fig:grainsizes}. Grains with small energy barriers reorient during a magnetization curve cycle. They can be divided into superparamagnetic grains\cite{Ehr11_drift} showing the smallest energy barriers (Class I) and coercivity mediating grains\cite{OGr10_york} (Class II). Only grains with higher energy barriers having a relaxation time longer than the measurement time are able to contribute to the EB field \heb \cite{OGr10_york}. The contribution of these thermally stable grains to the direction of \heb~is random, as long as they are not set by a magnetic field cooling process\cite{OGr10_york} or a similar procedure\cite{engel2005initialization,Har12_model}. They can be divided further into grains, which reorient during the field cooling process yielding a macroscopic contribution to \heb~(Class III) and the grains with the highest energy barrier (Class IV), which are even stable at the field cooling conditions.
\\In addition to the influence of the AF layer on the EB, the magnetic anisotropy of the F layer itself is another fundamental property defining the behavior of the system. For the characterization of EB samples the influences of the different magnetic anisotropy terms\cite{Cam05_aniso} have to be distinguished. Angular resolved measurements\cite{Amb97_angle} of \heb~and the coercive field \hc~(see figure \ref{fig:hys_schema}) may be used to disentangle the influences of the distributions of the different magnetic anisotropies\cite{Hof03_tailor,Cam05_aniso}, if these measurements are compared to numerical calculations, e.g., based on the model of Stoner and Wohlfarth\cite{Sto47_mod,Sto48_mod}. With this approach material constants and the mutual direction of different magnetic anisotropy contributions are visualized\cite{Jim09_noncoll}. 
\\In none of these numerical models, however, the relaxation time distribution\cite{OGr10_york} of the AF grains is considered. Although this may not be necessary at low temperature, where some of the fundamental experiments have been carried out, applications of the EB systems usually take place at room temperature (RT). A proper characterization, therefore, needs to be performed at RT, where thermal activation processes can not be neglected\cite{OGr10_york}. To reveal the impact of Class II grains the typical procedure of measuring magnetic easy axis hysteresis loops is not sufficient for EB systems possessing non negligible magnetic anisotropy of the F layer, since the two sources of coercivity can not be distinguished.
\\Our numerical calculations are based on the model of Stoner and Wohlfarth\cite{Sto47_mod,Sto48_mod}. Thermal effects\cite{Ful72_mod,OGr10_york} are introduced by taking into account a rotatable magnetic anisotropy term related to the grain size. Further, we apply angle-resolved magnetization curve measurements using the magnetooptic Kerr effect (MOKE) to determine \heb~and \hc~as a function of angle for a model system consisting of an \irmn/\cofe~bilayer. By comparing these results to the numerical model it is possible to reveal the magnetic properties of the system, including the effects arising from thermally unstable parts of the AF, by distinguishing the different sources of coercivity. Further we show the impact of the measurement time on \heb~and \hc~as a function of angle to prove the validity of the proposed model.

\section{Experimental}
\subsection{Sample preparation}
The EB bilayer system of \irmn$^{30~\mathrm{nm}}$/\cofe$^{15~\mathrm{nm}}$ was deposited on a naturally oxidized Si(100) substrate using rf-sputter deposition at room temperature with an applied in-plane magnetic field of 60~kA/m, where the base pressure was $10^{-6}$~mbar and the working pressure $2\cdot 10^{-2}$~mbar. A 50~nm Cu buffer layer was used to induce the (111) texture in the \irmn \cite{Ale08_text}. For the AF a layer thickness of 30~nm was chosen to enhance the AF grain volume delivering high thermal stability \cite{OGr10_york} with reduced thermal activation. A Si capping layer with a thickness of 20~nm was used to protect the EB bilayer from oxidation and for enhanced contrast in magnetooptic measurements \cite{Nak85_enhance}.
\\The EB system subsequently was annealed at 573.15~K for 60~min in an external in-plane magnetic field of 80~kA/m to maximize the macroscopic \heb. Afterwards the samples were cooled down in this field to room temperature at a rate of 5~K/min. 
\subsection{Magnetooptic Kerr effect measurements}
Samples were investigated by vectorial magnetooptic Kerr magnetometry in an extended MOKE setup similar to the one described in Ref.\cite{kuschel2011vectorial}. $p$-polarized light from a laser operating at a central wavelength of 632~nm was used to illuminate the sample. The reflected light was analyzed by a detector system yielding the reflectivity and the Kerr angle of the sample. The Kerr angle, in case of an in-plane magnetized sample, refers to the longitudinal Kerr effect and, therefore, is directly proportional to the magnetization component parallel to the applied magnetic field\cite{mag_domains}. The reflectivity of the sample yields direct proportionality to the transverse magnetization component. Normalizing both components with respect to the saturation magnetization allows reconstruction of the magnetization vector \cite{Vas00_pol}.
\\The sample is mounted on a rotatable sample stage in order to perform angular resolved measurements within the sample plane. Magnetization curves were obtained over an external magnetic field angle range of 360\gra~ with a resolution of 2\gra~applying an external magnetic field divided into 300~steps per branch with a maximum magnetic field of 80~kA/m. Due to enhanced magnetooptic effects arising from the silicon capping layer, no averaging of magnetization curves was necessary.
\\For the time dependent measurements one set of magnetization curve measurements was recorded for different measurement times $T_\mathrm{H}$ for one magnetization curve. In each set of measurements $T_\mathrm{H}$ was kept constant and the external magnetic field angle was varied in a range of 180\gra~with a resolution of 2\gra. $T_\mathrm{H}$ was selected between 17~seconds and 5~minutes. The sets were measured in a random order to make sure that the training effect is not the main reason for the observed changes in the magnetization curves.
\label{kerr}
\section{Model}
\begin{figure}[htbp]
\includegraphics{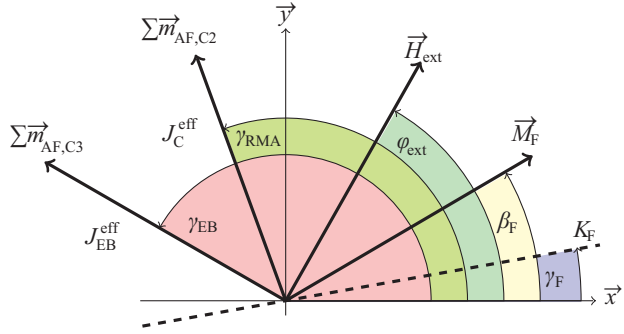}
\caption{\label{fig:ebmodel}Graphical illustration of the angles and vectors used in the proposed model in a polar coordinate system. $\vec{M}_\mathrm{F}$~is the magnetization vector of the F layer and \betf~its azimuth. \kf~is the energy density per unit area of the uniaxial magnetic anisotropy (UMA) of the F and \alf~the azimuth of its magnetic easy direction. \vhext~is the external magnetic field vector with the azimuth \phext. $\vec{m}_\mathrm{AF,C2}$ and $\vec{m}_\mathrm{AF,C3}$ are the surface magnetization moments associated to AF grains of Class II and III respectively. \alra~and \aleb~are the corresponding azimuths of the vector sums, which are connected to the rotatable and unidirectional magnetic anisotropies with the respective surface energy densities \jhc~and \jeb.}
\end{figure}

For the numerical calculations of the magnetization curves a Stoner-Wohlfarth-like model was used, where the magnetization of the F $\vec{M}$ was assumed to be uniform \cite{Sto47_mod}. This holds as long as the magnetization reversal of the systems occurs via coherent rotation. However, for magnetization curve measurements along the magnetic easy axis of the system the magnetization reversal takes place via nucleation and/or domain wall motion \cite{McC03_asym}, so deviations in this regime are expected. Due to strong shape anisotropy of thin film EB systems the magnetization was assumed to be parallel to the surface.
\\To determine the magnetization direction of the F layer a potential energy landscape is calculated as a function of \betf, which is the angle between $\vec{M}$ and the x-axis of the coordinate system (see figure \ref{fig:ebmodel}). The potential energy landscape per area $A$ is minimized with respect to \betf. For the case of more than one minimum the magnetization direction is derived with the perfect delay convention \cite{Nie91_mic}. Therefore, only that energetic states are taken into consideration, which are reachable via rotation from the starting point of the \betf~ variation without overcoming an energy barrier.
\\The potential energy landscape consists of several magnetic anisotropy terms which define the behavior of the system (see figure \ref{fig:ebmodel} for a graphical illustration) and the Zeeman term $E_{\mathrm{Z}}$, describing the interaction of the magnetic layer with the external magnetic field with
\begin{equation}
\label{eq_zee}
E_{\mathrm{Z}}/A=\mu_0\hext\msat\tf\cos{\left(\betf-\phext\right)}.
\end{equation}
Here, $\mu_0$ is the magnetic permeability in vacuum, \msat~the saturation magnetization of the F and \tf~its thickness, \hext~the strength of the external magnetic field and \phext~the angle describing its in-plane direction. 
\\The intrinsic magnetic anisotropy $E_{\mathrm{uni}}$ of the F layer was assumed to be uniaxial\cite{Kim06_ang}. This is valid, although the magneto-crystalline anisotropy of CoFe is biaxial \cite{Oks11_text,Kus12_cofe}, because the uniaxial magnetic anisotropy (UMA) induced by the field cooling process is dominant. This was also confirmed in preparatory investigations (not shown), where angle-resolved vector MOKE measurements with a pure \cofe~layer were performed. The energy area density of the UMA results in
\begin{equation}
\label{eq_uni}
E_{\mathrm{uni}}/A=\kf\tf \sin^2{\left(\betf-\alf\right)}.
\end{equation}
\kf~is the energy volume density constant of the UMA and \alf~the angle between the magnetic easy direction of the UMA and the x-axis of coordinate system.
\\To accurately quantify the influence of the AF, the material parameters of each individual AF grain which define the interaction with the F (volume, shape or magnetic anisotropy) have to be known. Here the real situation is approximated by using one average magnetic anisotropy term for each of the above mentioned classes of grains. At first there is no magnetic anisotropy term needed for grains of Class IV, although each of these grains contributes an individual unidirectional magnetic anisotropy to the F. Due to the statistical orientation of the individual unidirectional anisotropies the sum of all of these energy terms becomes zero within the used coherent rotation model. Thus, no energy term needs to be included for these grains. Superparamagnetic grains (Class I) are also neglected for the same reason. The important magnetic anisotropy terms are of grains of Class II and III, which are described as follows:
\\Class III grains are modeled by a cosine term as in the original model of Meiklejohn and Bean \cite{Mei56_mod}, but with a different interpretation. It is 
\begin{equation}
\label{eq_jeb}
E_{\mathrm{EB}}/A=-\jeb \cos{\left(\betf-\aleb\right)}.
\end{equation}
The effective exchange energy constant \jeb~sums up the interactions of all thermally stable AF grains\cite{Ehr05_mod} and \aleb~describes the average direction. In a system, where there is no statistical orientation of the magnetic moments in the AF before field cooling, \jeb~also takes the net contribution of that Class IV grains into account whose magnetic interactions are not compensated by other Class IV grains. In a system, where the individual contributions have different directions (e.g. IrMn with random in-plane magnetic easy axis distribution), the local exchange interaction constants between the individual grains and the F can be much higher \cite{Ste04_aniso}. Thus, a comparison between \jeb~and theoretical values for the exchange interaction of perfect interfaces usually fails.
\\The surface magnetization of Class II grains has its preferred direction close to the magnetization direction of the F layer and relaxes into its preferred state within the corresponding relaxation time. The resulting magnetic anisotropy contribution of these grains is, therefore, a rotatable magnetic anisotropy (RMA), which was described in several models before\cite{Sti98_mod,Ges02_rot}, although not all of these models are connected to polycrystallinity\cite{Rad08_rev}. The RMA in these models is either considered as an energy term favoring the apparent magnetization direction\cite{Sti98_mod} or the actual axis of the external magnetic field\cite{Ges02_rot}. It is not considered in detail, however, on which timescale the reorientation of the RMA takes place. For thermally unstable grains, this timescale is the relaxation time which may be distributed over several orders of magnitude. For a typical grain size distribution \cite{OGr10_york} there is coexistence of grains which reorientate almost immediately and of grains which do not remagnetize before a magnetization curve measurement has ended. The magnetic anisotropy $E_{\mathrm{RMA}}$ connected to Class II grains, therefore, energetically favors the former magnetization direction of the F \alra~at the time $t-\taua$, where $t$ is the time and \taua~the average relaxation time of the individual grains. The energy area density for the RMA is
\begin{align}
\label{eq_rot}
E_{\mathrm{RMA}}/A&=-\jhc \cos{\left(\betf-\alra\right)}
\\\mathrm{with}~ \alra &= \betf(t-\taua).
\end{align}
The energy surface density of this macroscopic magnetic anisotropy \jhc~arises from the sum of all individual contributions of Class II grains. Modeled in this way this energy term requires a random magnetic easy axis distribution as it is typical for IrMn\cite{Ste04_aniso}. If this is not the case the direction of the RMA is biased and the exact distribution needs to be taken into account \cite{Har12_model}. 
\\Summing up all of the above mentioned energy contributions the total surface energy density writes as
\begin{equation}
E_1=E_{\mathrm{Z}}+E_{\mathrm{uni}}+E_{\mathrm{EB}}+E_{\mathrm{RMA}}.
\label{eq_mod}
\end{equation}
Note, that due to equation (\ref{relaxtime}) \jhc, \taua, and \jeb~depend on temperature and measurement time, with \jeb~also being affected by the field cooling conditions \cite{OGr10_york}; i.e. a comparison of experimental data is only useful if the measurement conditions are kept constant. 
\section{Results}
\subsection{Impact of different magnetic anisotropies}
\begin{figure}[htbp]
\includegraphics{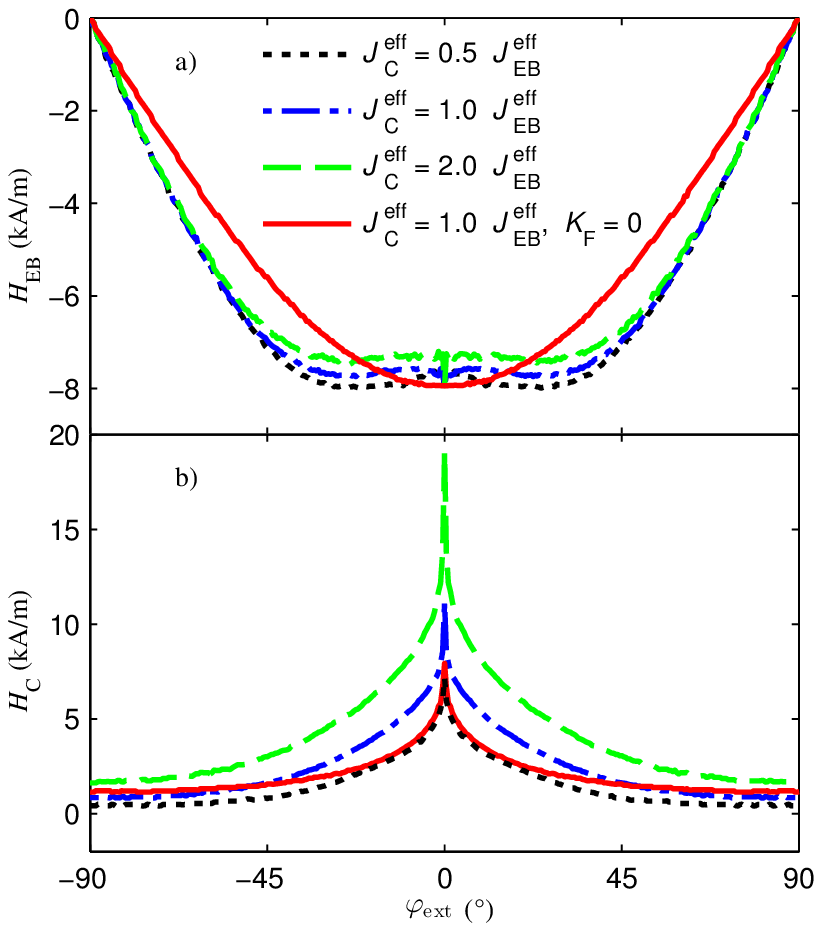}
\caption{\label{fig:hebhc}(a) \heb~and (b) \hc~as a function of angle calculated using equation (\ref{eq_mod}) for different \jhc. Material parameters used for the RMA were $\jhc=0.1$~mJ/m$^{2}$ (red line), $\jhc=0.05$~mJ/m$^{2}$ (black dotted), $\jhc=0.1$~mJ/m$^{2}$ (blue dash-dotted) and $\jhc=0.2$~mJ/m$^{2}$ (green dashed). For the UMA $\kf=0$~J/m$^{3}$ (red line) and $\kf=2000$~J/m$^{3}$ (others) was used. Other material parameters can be found in table \ref{tab_parvals}.}
\end{figure}

\begin{figure}[htbp]
\includegraphics{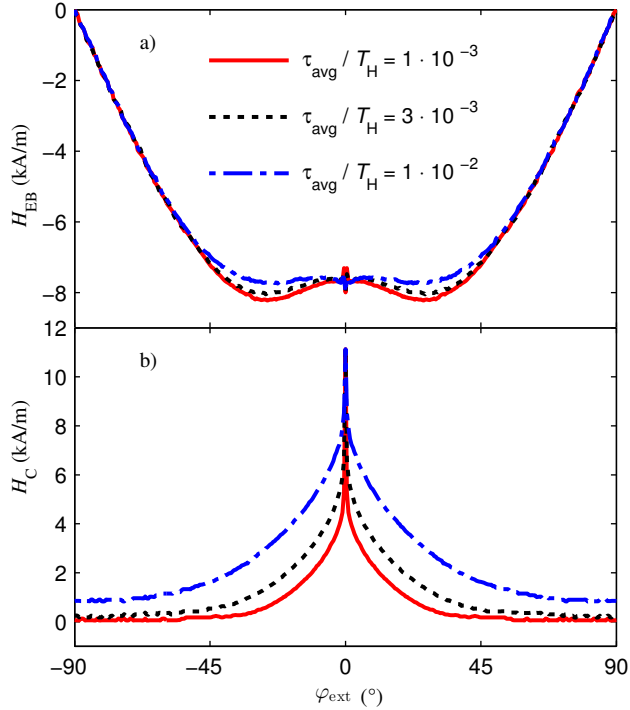}
\caption{\label{fig:stepwidth}(a) \heb~and (b) \hc~as a function of angle calculated using equation (\ref{eq_mod}) for different \taua. Material parameters used for the RMA were $\taua/T_\mathrm{H}=10^{-3}$ (red line), $\taua/T_\mathrm{H}=3\cdot 10^{-2}$ (black dotted), $\taua/T_\mathrm{H}=1\cdot 10^{-2}$ (blue dash-dotted). Other material parameters can be found in table \ref{tab_parvals}.}
\end{figure}

Equation (\ref{eq_mod}) yields two sources of coercivity: $E_{\mathrm{uni}}$ and $E_{\mathrm{RMA}}$. It is possible to distinguish between these two different sources by recording magnetization curves at different in-plane angles of the magnetic field axis. To show the impact of the different magnetic anisotropies on the angular resolved EB and coercive fields \hebt~and \hct, numerical simulations using the proposed model were performed by varying one source of coercivity at a time. Table \ref{tab_parvals} shows the parameters used in the calculations, where \taua~was defined in fractions of the time $T_\mathrm{H}$ needed for one magnetization curve.
 \begin{table}
 \caption{\label{tab_parvals} Parameters used for the numerical calculations in figures \ref{fig:fm_aniso}, \ref{fig:hebhc} and \ref{fig:stepwidth}. The calculations were carried out using steps of 0.5\gra~for \betf~ with 1000 field steps per magnetization curve.}
 \begin{ruledtabular}
 \begin{tabular}{c|c|c|c}
 \textbf{Material Constant} & \textbf{Var \kf}& \textbf{Var \jhc}& \textbf{Var \taua}\\
\hline
\kf~(J/m$^{3}$)&500 - 4000&0; 2000&2000  \\
\jhc~(mJ/m$^{2}$)&0&0.05 - 0.2&0.1 \\
\taua$/T_\mathrm{H}$&unused&$10^{-2}$&$10^{-3}$ - $10^{-2}$\\
\hline
\jeb~(mJ/m$^{2}$)&\multicolumn{3}{c}{0.1}\\
\tf~(nm)&\multicolumn{3}{c}{10}\\
\msat~(kA/m)&\multicolumn{3}{c}{1000}\\
\alf~(\gra) &\multicolumn{3}{c}{0}\\
\aleb~(\gra)&\multicolumn{3}{c}{0}\\
\hext~(kA/m) &\multicolumn{3}{c}{$\pm 40$}
 \end{tabular}
 \end{ruledtabular}
 \end{table}
\begin{figure}[htbp]
\includegraphics{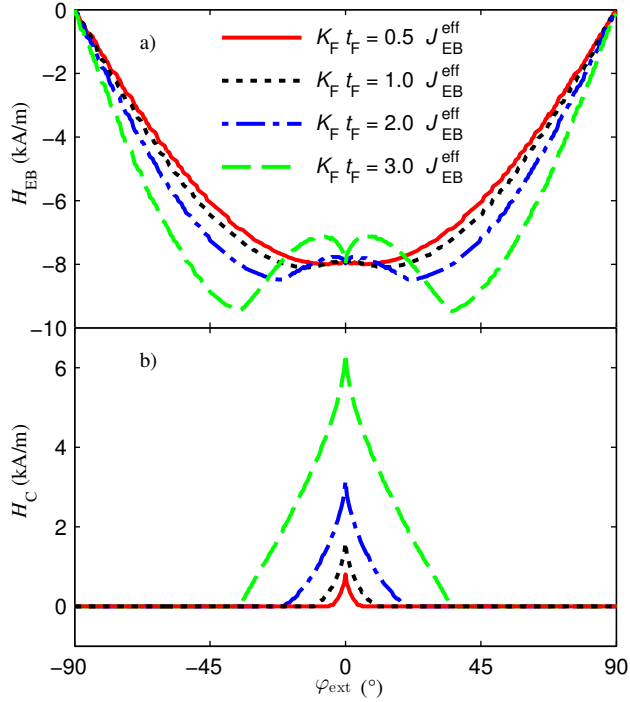}
\caption{\label{fig:fm_aniso}(a) \heb~and (b) \hc~as a function of angle calculated using equation (\ref{eq_mod}) for different \kf. Material parameters used for the UMA were $\kf=500$~J/m$^{3}$ (red line), $\kf=1000$~J/m$^{3}$ (black dotted), $\kf=2000$~J/m$^{3}$ (blue dash-dotted) and $\kf=4000$~J/m$^{3}$ (green dashed). Other material parameters can be found in table \ref{tab_parvals}.}
\end{figure}
\\From figure \ref{fig:fm_aniso} it is evident, that the magnetic anisotropy of the F not only increases \hc, but also has a big influence on the shape of \heb~as a function of \phext. This is in accordance to previous calculations \cite{Cam05_aniso,Bai10_angle}, where the angular dependence of EB and coercive fields was calculated for the classical model of Meiklejohn and Bean without using a RMA term. The most prominent characteristics are the triangular shape of the coercivity with a broadened base for higher UMA of the F and the maximum of \hebt~shifting away from the magnetic easy axis with increased magnetic anisotropy in the F. The principle shape of both characteristics is solely defined by the strengths of the exchange anisotropy and the F anisotropy.
\\However, the influence of the second source of coercivity, i.e. the RMA, is different. From figure \ref{fig:hebhc} it is obvious that the general shape of \hebt~is almost not altered by the relative strength of the RMA. The dependence of \hebt~on the strength of the RMA decreases for EB systems possessing smaller UMA and vanishes for systems without F anisotropy. The shape of the coercivity mediated by $E_{\mathrm{RMA}}$ in general is not triangular but Lorentzian-like. For systems, which have RMA and UMA, the shape of the coercivity in first order is a superposition of both contributions.
\\From figure \ref{fig:stepwidth} it is evident, that \heb~and \hct~not only depend on the strength of the RMA but also on \taua. For an increased average relaxation time of the Class II grains the plateau of \hebt~close to the magnetic easy axis of the system flattens and the peak in \hct~becomes broader. For long \taua~the minimum of \hct~is increased leading to a non vanishing coercive field for \phext~far away from the magnetic easy axis of the system.
\\Looking at the angular dependencies of \hebt~and \hct, respectively, there is a clear difference in the influence of the two sources of coercivity. While \hebt~is strongly affected by the UMA, \hct~has different shapes for the two magnetic anisotropies mediating coercivity. Therefore, it is possible to determine the dominant magnetic anisotropy being responsible for the coercivity by measuring \hc~as a function of \phext.
\subsection{Comparison with experiment}
\begin{figure}[htbp]
\includegraphics{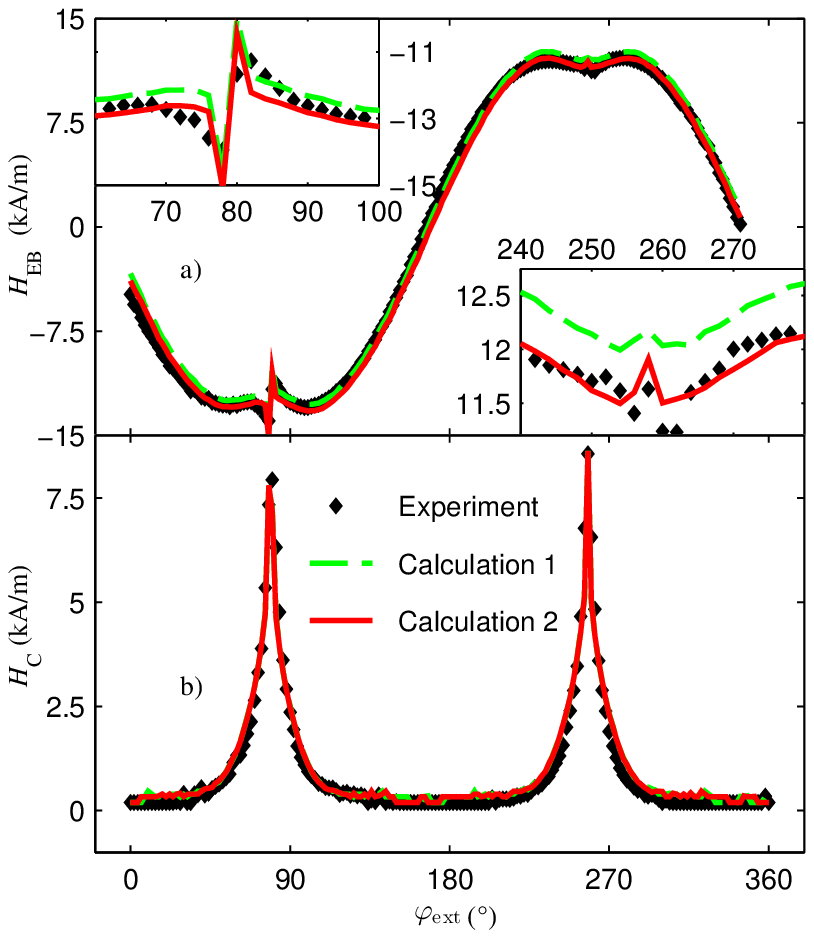}
\caption{\label{fig:result}(a) \heb~and (b) \hc~ as a function of angle (black symbols) of a system of Cu$^{50~\mathrm{nm}}$/\irmn$^{30~\mathrm{nm}}$/\cofe$^{15~\mathrm{nm}}$/Si$^{20~\mathrm{nm}}$. The dashed line (green) corresponds to numerical calculations based on equation (\ref{eq_mod}) using a resolution of 0.5\gra~for \betf. Material parameters used can be found in table \ref{tab_simdata}. The red line corresponds to a numerical calculation based on equation (\ref{eq_mod2}) using the same parameters. For \hc~there is almost no difference for both calculations.}
\end{figure}

The model was tested by comparing it to experimental data of \hebt~and \hct, obtained by vectorial magnetooptic Kerr magnetometry (figure \ref{fig:result}). The experimentally determined relation of \hebt~is similar to one of the calculations of figure \ref{fig:fm_aniso}, indicating that the system possesses the assumed UMA (here with the magnetic easy axis at $\alf\approx 80\gra$) of the Co$_{70}$Fe$_{30}$ layer. In contrast to the triangular shape of \hct~for the simulated system having UMA as the only source of coercivity, the experimentally detected shape of \hct~is curved. Therefore, elements of both coercivity mediating magnetic anisotropies are visible in the experiment.
\\Additionally, there is a Fano-like structure\cite{fano1961effects} in \hebt~for measurements along the magnetic easy direction at an angle of about 80\gra~observable, which does not appear for the other magnetic easy direction around 260\gra~measured later. Such a curvature results, when \aleb~and \alf~are not aligned parallel\cite{Jim11_aniso}, which is a result of a misalignment between the external magnetic fields during sample deposition and field cooling. In the present case this misalignment is very small and vanishes during the measurement. This may be connected to training effects, especially because it is well known that thermal activation can change the direction of EB\cite{vdH98_therm}.
\\A fit of the material properties by the model in equation \ref{eq_mod} to the experimental data yields a numerically calculated dependency of \hebt~and \hct, which agrees almost perfectly with the experimental values (green line in figure \ref{fig:result}). The normalized root-mean-square deviation $\sigma$ as a matching factor yields $\sigma_\mathrm{EB}=2.4~\%$ for the EB and $\sigma_\mathrm{C}=3.6~\%$ for the coercive field relation. Due to the change of \aleb~indicated by the curvature, different values for \aleb~where assumed for $\phext<180\gra$ and $\phext>180\gra$.
\\The calculated values for \hebt, however, are slightly bigger, because the absolute values in the experiment are not equal for the positive and the negative part of the relation. The discrepancy is not connected to the thermal training effect \cite{Sch68_train,Bin04_train}, because the difference is reproducible in repetitive measurements. We, therefore, connect this phenomenon with the measurement procedure: Before each magnetization curve is recorded a calibration of the detector system was performed, allowing a larger number of AF grains to relax into the energetic state favored by the apparent magnetization direction. Therefore, the number of grains which contribute to the magnetization curve shift is increased (decreased), when the apparent magnetization direction of the F is parallel (antiparallel) to the magnetic easy direction of the unidirectional magnetic anisotropy. This effect can be accounted for by an additional magnetic anisotropy term $E_{\mathrm{add}}$, which energetically favors the direction of the initial magnetic field $\varphi_{\mathrm{ini}}$ of each magnetization curve leading to the total energy density
\begin{equation}
E_2 =E_1+E_{\mathrm{add}},~\mathrm{with}~E_{\mathrm{add}}=-J_{\mathrm{eff}}^{\mathrm{add}}\cos{\left(\beta_{\mathrm{F}}-\varphi_{\mathrm{ini}}\right)}
\label{eq_mod2}
\end{equation}
\\Here, \jadd~is the effective exchange interaction connected to the additional grains relaxing at the beginning of each magnetization curve. Using equation (\ref{eq_mod2}) for the calculation (red line in figure \ref{fig:result}), a very small value for \jadd~reduces the deviations further to $\sigma_\mathrm{EB}=1.9\%$ and $\sigma_\mathrm{C}=3.6\%$.
 \begin{table}
 \caption{\label{tab_simdata} Material constants obtained by fitting the numerical calculations to the experimental data in figure \ref{fig:result}. The error corresponds to the uncertainty in the fit constants introduced by a variation of the starting conditions by a factor of 3.}
 \begin{ruledtabular}
 \begin{tabular}{cc}
 \textbf{Material Constant} & \textbf{Used value}\\
\hline
\kf 		&$(3300\pm300)$ J/m$^{3}$ \footnote{fitted to experiment}\\
\tf 		&15 nm \footnote{value given by experiment}\\
\msat 	&1230 kA/m \footnote{measured by superconducting quantum interference device}\\
\jeb		&$(0.285\pm0.03)$~mJ/m$^{2}$ \footnotemark[1]\\
\jhc		&$(0.15\pm0.2)$~mJ/m$^{2}$ \footnotemark[1]\\
\jadd		&$(0.011\pm0.002)$~mJ/m$^{2}$ \footnotemark[1]$^,$\footnote{used only for calculations based on equation (\ref{eq_mod2})}\\
\taua		&$(300\pm100)$~ms\footnotemark[1]\\
$\aleb-\alf(\phext < 180\gra)$ & $(2.5\pm1)$\gra~\footnotemark[1]$^,$\footnote{experimental offset depends on sample position}\\
$\aleb-\alf(\phext > 180\gra)$ & $(0\pm1)$\gra~\footnotemark[1]$^,$\footnotemark[5]
 \end{tabular}
 \end{ruledtabular}
 \end{table}
\\It is possible to use the model for revealing the important magnetic material properties of EB layer systems including effects related to the micro magnetic fine structure of the AF. We believe the error of most of the received material properties to be smaller than 10~\%, because even strong variations in the starting conditions of the fit by a factor of 3 results in almost the same material constants. Especially \jeb~and the misalignment between \aleb~and \alf~can be detected with great precision. The calculations are not so sensitive on \taua, where the uncertainty is about 30~\%. This is not unexpected, because the relaxation times of the individual AF grains differ by several orders of magnitude \cite{Ful72_mod,Ehr05_mod}.

\subsection{Influence of the measurement time}
\begin{figure}[htbp]
\includegraphics{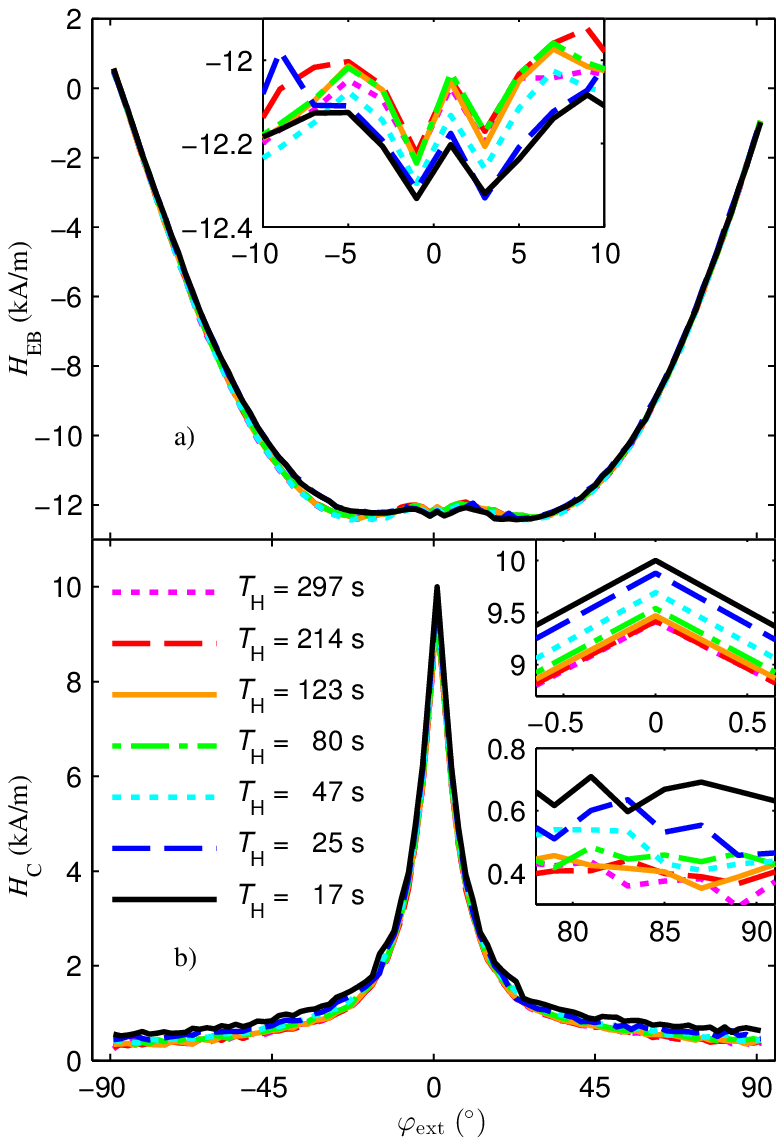}
\caption{\label{fig:time_dep_full}(a) \heb~and (b) \hc~ as a function of angle of a system of Cu$^{50~\mathrm{nm}}$/\irmn$^{30~\mathrm{nm}}$/\cofe$^{15~\mathrm{nm}}$/Si$^{20~\mathrm{nm}}$ for different measurement times of the magnetization curves $T_\mathrm{H}$.}
\end{figure}

\begin{figure}[htbp]
\includegraphics{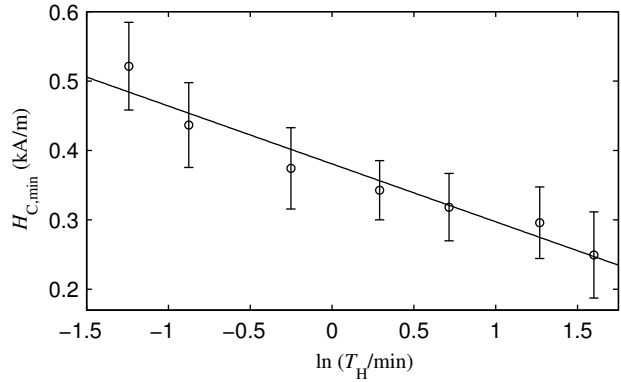}
\caption{\label{fig:time_dep_qual}Minimum of the relation \hct~for different $T_\mathrm{H}$. The error corresponds to the uncertainty in the determination of \hc. The line is a linear fit serving as a guide to the eye.}
\end{figure}

The proposed model and its precursors strongly focus on the impact of the relaxation times of the individual AF grains.\cite{Ful72_mod,Ehr05_mod,OGr10_york} Thus, it is very important to take care of the experimental timescales, because they define how to classify the AF grains into the four different categories, i.e. the position of the borderlines in figure \ref{fig:grainsizes}. To show the impact of the experimental timescales and to verify our approach on how to design the RMA, \hebt~and \hct~were measured in dependence of $T_\mathrm{H}$. The impact of the available timescales on this relations is undeniably small as depicted in figure \ref{fig:time_dep_full}.
Nevertheless, it is possible to see the impact of the measurement time by looking at the extreme values. The impact of the measurement time is similar to the impact of temperature in general as the underlying mechanisms can be described by the N\'{e}el-Arrhenius law. Therefore, we do not want to focus on the behavior of the absolute maximum values of \heb~and \hc, which were discussed several times before and are well described by the polycrystalline model.\cite{Ful72_mod,OGr10_york} The situation is different for the minimal values \hcmin~of the relation \hct~(see figure \ref{fig:time_dep_qual}), which correspond to the heavy axis magnetization curves.
In first order \hcmin~as a function of $T_\mathrm{H}$ shows a logarithmic decrease (linear decrease in the logarithmic presentation). In the calculations in figure \ref{fig:stepwidth} it was shown that \hcmin~is nonzero if a RMA is used under consideration of the average relaxation time of the Class II grains. Further, \hcmin~was increased for higher fractions of \taua/$T_\mathrm{H}$. Despite the fact that \taua~should strongly depend on $T_\mathrm{H}$ because changing the timescale dramatically changes the thermal stability of the AF grains, it is very reasonable that the fraction \taua/$T_\mathrm{H}$ is decreased for longer $T_\mathrm{H}$. Thus, the increase of \hcmin~for faster measurements is expected by theory and, therefore, is another argument for the validity of the proposed model.

\section{Conclusion}
In this study, we have shown a model based on the concepts of Stoner and Wohlfarth, which allows numerical calculations of \heb~and \hc~ in dependence of the external magnetic field angle including the thermal instabilities of the polycrystalline AF layer in a fast and simple approach. We were able to show, that the two main sources of coercivity, namely the F magnetic anisotropy and the RMA resulting from the exchange interaction of thermally unstable grains, can be disentangled via the angular dependency of \heb~and \hc. Adjusting the calculations to the experimental data shows excellent agreement. For a Si/Cu$^{50~\mathrm{nm}}$/\irmn$^{30~\mathrm{nm}}$/\cofe$^{15~\mathrm{nm}}$/Si$^{20~\mathrm{nm}}$ system F magnetic anisotropy with $\kf=(3300\pm300)~\mathrm{J~m^{-3}}$ and uniaxial anisotropy with $\jeb=(0.285\pm0.03)~\mathrm{J~m^{-2}}$ was determined. For the RMA an energy density of $\jhc=(0.15\pm0.2)~\mathrm{J~m^{-2}}$ and an average relaxation time of $\taua=(300\pm100)~\mathrm{ms}$ was found. Further, we were able to prove our model as it predicts the evolution of \hebt~and \hct~when the measurements are performed on different time scales.
\\We suggest using this technique for sample characterizations, for example to study the influence of ion bombardment on EB samples, when a complete characterization of a sample including its grain size distribution and temperature dependency is too intricate. 

\section{Acknowledgments}
NDM thanks the University of Kassel for the Universit\"at Kassel Promotionsstipendium.
\bibliography{vmoke}

\end{document}